\documentclass{article}

\usepackage{arxiv}

\usepackage[utf8]{inputenc} % allow utf-8 input
\usepackage[T1]{fontenc}    % use 8-bit T1 fonts
\usepackage{hyperref}       % hyperlinks
\usepackage{url}            % simple URL typesetting
\usepackage{booktabs}       % professional-quality tables
\usepackage{amsfonts}       % blackboard math symbols
\usepackage{nicefrac}       % compact symbols for 1/2, etc.
\usepackage{microtype}      % microtypography
\usepackage{lipsum}
\usepackage{amsmath}
\usepackage{subfig}
\usepackage{graphicx}
\usepackage{url}

\title{Reactive Liquid: Optimized Liquid Architecture for Elastic and Resilient Distributed Data Processing}

\author{
  Seyed Esmaeil Mirvakili\\
  Department of Computer Engineering\\
  Sharif University of Technology\\
  Tehran, Iran \\
  \texttt{si.mirvakili95@student.sharif.edu} \\
   \And
 MohammadAmin Fazli \\
  Department of Computer Engineering\\
  Sharif University of Technology\\
  Tehran, Iran \\
  \texttt{fazli@sharif.edu} \\
  \And
 Jafar Habibi \\
  Department of Computer Engineering\\
 Sharif University of Technology\\
  Tehran, Iran \\
  \texttt{jhabibi@sharif.edu} \\
}

\begin{document}
\maketitle

\begin{abstract}
Today's most prominent IT companies are built on the extraction of insight from data, and data processing has become crucial in data-intensive businesses. Nevertheless, the size of data which should be processed is growing significantly fast. The pace of the data growing has changed the nature of data processing. Today, data-intensive industries demand highly scalable and fault tolerant data processing architectures which can handle the massive amount of data. In this paper, we presented a distributed architecture for elastic and resilient data processing based on the Liquid which is a nearline and offline big data architecture. We used the Reactive Manifesto to design the architecture highly reactive to workload changes and failures. We evaluate our architecture by drawing some numerical comparisons between our architecture prototype and the Liquid prototype. The performed evaluation shows that our architecture can be more scalable against workload and more resilient against failures than the Liquid architecture is.
\end{abstract}

% keywords can be removed

\section{Introduction}\label{sec:introduction}

Not only is data a fundamental value in business today but also many companies such as Google and Facebook are built on it. As a result, data processing is the primary concern of these companies. David Reinsel et al. \cite{Reinsel2017} have forecasted that size of the global datasphere, which is all data created and captured and replicated on earth, is going to reach 160 zettabytes by 2025. Hence, the companies are trying to improve their processing power to fit their business for such massive data. Distributed computing is one of the best solutions to increase the processing power.

Distributed systems have attracted many IT companies' attention in the last decades. This concept offers higher processing power by distributing the works among multiple decentralized components in a network. Big data companies have employed distributed technologies and frameworks such as HDFS (Hadoop Distributed File System) and MapReduce to handle the massive volume of data. HDFS, the file system component of Hadoop, provides fault tolerant data storage distributed on clusters \cite{Shvachko2010}.

Distributed systems are exposed to a broad range of network, hardware, and software failures. Hence, resiliency is of crucial importance in distributed processing. Resilient systems should not only be fault tolerant but also heal themselves and recover their original state. Moreover, distributed systems may be overloaded with the flood of users' requests. Thus, the system should be scalable to such an immense amount of workload, yet scalability alone cannot guarantee the system health. A system should scale itself automatically based on system status. In fact, the system should scale on demand, which is called elasticity.

We present an elastic and resilient architecture named Reactive Liquid for distributed data processing to confront distributed data processing challenges. The architecture which this paper proposes is designed based on the Liquid architecture, a nearline data integration architecture, since the Liquid is one of the best big data processing architectures which can fit the context of elastic and resilient distributed data processing. Nevertheless, the Liquid has some limitations to be a perfect solution to this problem. The most burdensome limitation of the Liquid architecture is that the scalability of the processing layer is limited to the messaging layer. The Reactive Liquid surmounts the shortcomings of the Liquid architecture by separating the processing layer from the messaging layer and makes it highly qualified for elastic and resilient distributed data processing.
The rest of the article is organized as follows.

Section 2 examines renowned big data architectures and the Reactive Manifesto as guidelines for constructing elastic and resilient systems. Section 3 elaborates on improving the Liquid architecture, to be the basis of our design, and designing the proposed architecture. Section 4 presents the evaluation results of the presented architecture. Section 5 summarizes the main conclusions of our work and introduces a future work.

\section{Background}\label{sec:background}
One of the most widely used data processing architecture is Lambda architecture which was proposed by Nathan Marz and his colleagues \cite{Marz2013}. Data immutability which states that only new data can append to the system and old data should never change is the mainstay of Lambda architecture. As a result, not only will the system tolerate the human errors but also data indexing and querying will be more straightforward. This architecture separates the data processing into online and offline isolated layers. Input data will be sent to both layers in order that the same processing logic can be executed on the data in online and offline environments. Then the results will be sent to another layer named Serving Layer to enable the backend systems to run their queries on them.
The drawback of the Lambda architecture is that developers should maintain two separate layers which do the same processing. Separation of these two layers can complicate the system integration and maintenance, and make the system delicate.

Jay Kreps has solved the problem of maintenance in Lambda architecture and proposed an architecture named Kappa \cite{Kreps2014}. In Kappa architecture, real-time and the batch layers are unified as a single stream processing layer. In this layer, multiple jobs are processing the input data. When reprocessing of the data is needed, a new job starts alongside the others. The new job stores the results in the serving layer for the backend systems. Kappa architecture is unable to set up an incremental processing job and create a processing pipeline. A processing pipeline is a set of multiple jobs connected in series, where the output of one job is the input of the next job. Incremental processing makes data processing faster and more efficient.

\begin{figure}[!t]
\begin{center}
\includegraphics[width=3in]{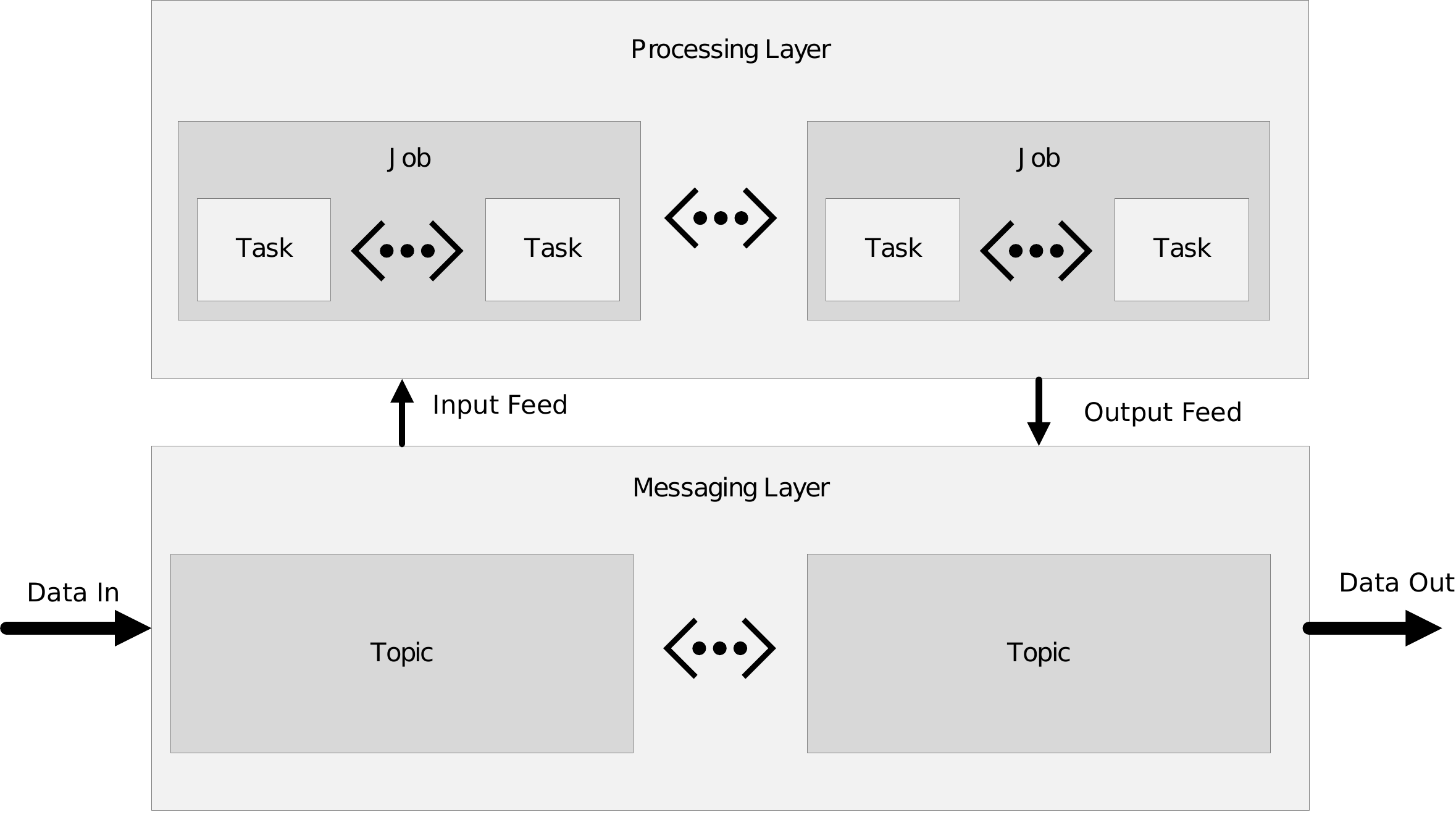}
\end{center}
\caption{The Liquid architecture consists of the Processing layer and the Messaging layer.}
\label{liquid_arch}
\end{figure}

Fernandez et al. \cite{Fernandez2015} proposed Liquid architecture which is a nearline data integration stack supporting near real-time and batch analytics. This architecture is shown in  \figurename~\ref{liquid_arch}. Incremental processing, low latency, and high availability are features of the Liquid architecture. The Liquid architecture separates the processing and data storing as two isolated layers, so it brings separation of concerns and makes the deployment and distribution of each layer easier. This architecture was deployed on more than 400 machines at LinkedIn to perform data integration and adaptation for back-end and front-end systems.

The layers of Liquid are:

\begin{enumerate}
\item Messaging Layer is a scalable distributed topic-based publish/subscribe communication platform which provides data access for the processing layer and backend systems.
\item Processing Layer consists of some jobs which execute data processing on messages from an input topic and publish the results as messages to an output topic. Every job is divided into some tasks which process a partition of the input topic to provide parallel processing.
\end{enumerate}

The jobs are isolated by the messaging layer, which provides loosely coupled components. Furthermore, the messaging layer is highly scalable and distributed. Hence, the Liquid architecture can fit the elastic and resilient distributed data processing whereas it has some limitations which are discussed in the next section.

The reactive manifesto \cite{Boner2014} offers some guideline to build reactive systems which should have four characteristics as follows:

\subsection{Message-Driven}
A reactive system should use asynchronous message-passing to bring loose coupling, isolation, and location transparency. In a reactive system, every communication between components is established by messages, even the failures. Being message-driven is a mean to achieve load balancing, elasticity, and flow control by monitoring the message queues between components. Moreover, asynchronous location transparent message-passing make it possible to distribute the system across a cluster without changing any part of the system. Some well-known asynchronous messaging technologies are RabbitMQ \cite{Videla2012} which is an open-source message broker middleware written in Erlang, Apache ActiveMQ \cite{Snyder2011} which is a high-available; reliable; scalable; and secure open-source message-oriented middleware(MOM) for enterprise messaging, Akka toolkit \cite{Roestenburg2015} which is an open-source JVM-based actor model toolkit for facilitating the construction of concurrent and distributed applications.

\subsection{Resilient}
A reactive system should stay responsive in the face of failure. The resiliency can be achieved by replication, containment, isolation, and delegation.

By replicating a component, we can ensure that the component will be available in case of a failure. There are several patterns for replication such as Active-Passive Replication, Multiple-Master Replication, and Active-Active Replication \cite{Kuhn2017}.\\
Containment means that failures must be contained in one component and not spread to other components. Component containment can be achieved by some patterns such as The Circuit Breaker \cite{Kuhn2017} which protects the users of a component in case of a prolonged failure condition by breaking the connection.

Isolation means that failure and recovery of a component must not affect the other components. Asynchronous message-passing communication establishes a boundary between components and isolates them so that a message-driven system can ensure the isolation of the components. Furthermore, the functionalities of a system must be isolated. In other words, a failure in functionality should not affect the others. Functionality isolation can be achieved by applying the Single Responsibility Principle \cite{Martin2002} on component-scale. This principle is an object-oriented principle which states that a class should have only one reason to change. In other words, a class should have only one responsibility. The Simple Component Pattern \cite{Kuhn2017} is a reactive design pattern which is derived from the single responsibility principle. This pattern states that a component should do only one responsibility, but do it thoroughly. By applying this pattern, every cohesive responsibility in the system will be placed in one component. 

Delegation means that the responsibility of recovering a failed component will be delegated to a healthy component called Supervisor so that users of the failed component do not have to deal with it. Component recovery consists of two stages. First, the supervisor detects a failure. Second, the supervisor tries to recover the failed component. There are some mechanisms such as Heartbeat \cite{Aguilera1997} and the $\varphi$ Accrual Failure Detector \cite{Hayashibara2004} to detect the failures. A component recovery should be performed out of the failed component. On the contrary, resolving the issues of a failed component from outside is nearly impossible. Hence, we need a simpler mechanism like the Let-It-Crash pattern \cite{Kuhn2017} to recover the failed component. This pattern states that it is preferable to restart a failed component instead of internal failure handling. Based on this pattern, the supervisor should restart the failed component in case of failure detection. In restarting, a stateful component should retrieve its last available state, so every stateful component should persist its state.

\subsection{Elastic}
A reactive system should react to changes in workload by increasing or decreasing the resources and stay responsive under these changes. Hence, reactive systems should not have any contention points or central bottlenecks. By elimination of contention points and central bottlenecks from the system, we can shard and replicate the components of the system and distribute. Furthermore, we can distribute and balance the inputs among the components. Reactive systems measure the performance of components continuously to enable a predictive and reactive scaling. By live performance monitoring, the system can detect or predict changes in workload. Then the system can react to these changes by replicating or sharding the components.

An elastic system should be able to replicate the components fast and automatically without human intervention. Moreover, new component instances which are created by replicating or sharding should be location transparent to the system. In fact, instantiation and deployment of the new instances in physical infrastructures should be separate from the system. Furthermore, the system should provide a fair load balancing for component instances. Some technologies and computation models exist which can provide us with elastic development and design, such as Actor Model \cite{DeKoster2016} which is a mathematical model of concurrent computation. Actor model offers isolation, message-driven communication, and flexibility. There are some actor model implementations widely available such as Akka toolkit \cite{Roestenburg2015}, Erlang programming language \cite{Armstrong1993} which was the first language in the industry adopting the actor model as its concurrency model.

Containerization technologies \cite{Soltesz2007} such as Docker \cite{Docker} and Rocket \cite{Rocket} are great tools to develop and deploy elastic systems. The containerization provides virtualized operating systems like traditional virtualization with the ability to run multiple instances of a system on one machine. By this technology, new instances of the components can be created automatically and immediately on demand. For example, the system can create new instances of a component which is under pressure. Furthermore, there are some technologies to achieve an elastic and resilient cluster of containers such as Kubernetes \cite{Kubernetes} which provides automatic deployment, scaling, and management of containerized applications.

Infrastructure as a Service (IaaS) \cite{Bhardwaj2010} and Platform as a Service (PaaS) \cite{Lv2010} are other options to develop an elastic distributed system. IaaS delivers the hardware and associated software, like operating system, as a service. By IaaS, resources can be allocated to the system by demand. OpenStack \cite{OpenStack} is an open-source cloud computing platform which is mostly deployed as IaaS. OpenStack provides a dashboard and an API to control large pools of resources such as computation, storage, and networking. PaaS provides a platform which enables the developer and operations to develop, run, and manage applications on demand and automatically. Moreover, PaaS hides the complexity of building and maintaining the infrastructure from customers. Amazon AWS \cite{Amazon_AWS} is one of the famous PaaS providers which provides on-demand cloud computing platforms.

\subsection{Responsive}
A reactive system should not only respond promptly but also detect the problems as quickly as possible. Moreover, a reactive system must be responsive in every situation such as facing failure or high workload. Moreover, the responsiveness of the system is dependent on resiliency and elasticity of the system.

\section{Architecture And Design}\label{sec:architecture}
\subsection{Improving the Liquid}
The separation of data and processing layers in Liquid architecture fits it for distributed systems. Furthermore, this separation may cause the isolation of jobs. In incremental processing, each job is separated by the messaging layer from other jobs so that every message and connection goes through the messaging layer. Separation of jobs helps us to break down a processing logic to some distributed jobs and scatter them across the network. The Liquid architecture has eliminated the problems in Lambda and Kappa, which are complicated maintenance and lacking an incremental processing solution, but there are still some problems in Liquid which are presented in what follows.

\begin{figure}[!t]
\begin{center}
\includegraphics[width=3in]{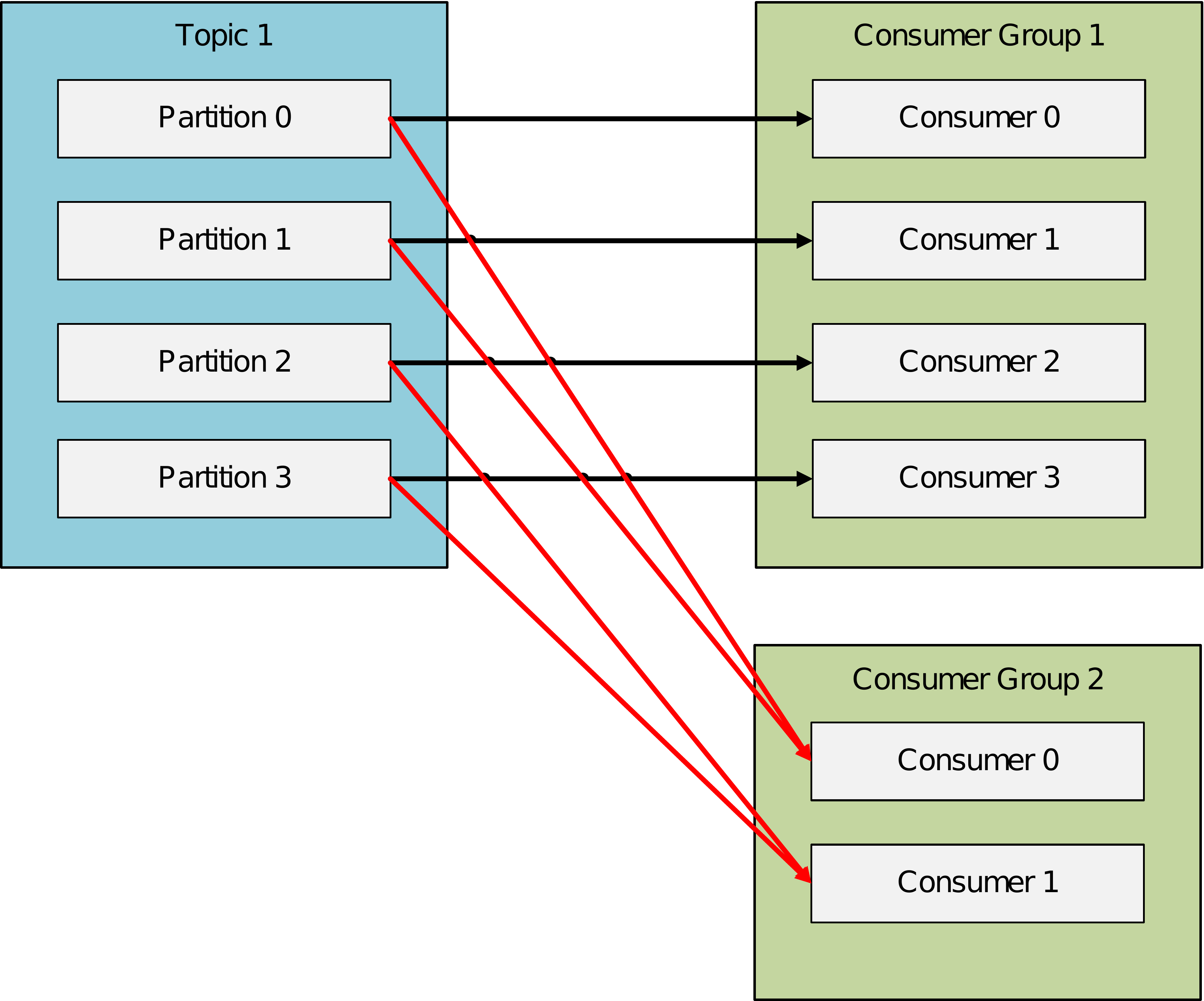}
\end{center}
\caption{Mapping Kafka partitions among consumers of consumer groups.}
\label{kafka_consumer}
\end{figure}

The messaging layer of Liquid uses Apache Kafka \cite{Garg2013} which is a distributed publish/subscribe messaging system. In Apache Kafka, data is encapsulated as messages which are stored in topics. Producers can publish messages on topics and consumers can subscribe on a topic and wait for published messages. Each topic consists of one or more partitions. In fact, messages of a topic are distributed across these partitions. Each consumer is labeled with a consumer group name, and all consumer instances with the same consumer group name are in the same consumer group. As shown in \figurename~\ref{kafka_consumer}, every partition of a topic can be consumed by only one of the consumers within a consumer group so that the number of working consumers of a consumer group is limited to the number of partitions of the topic from which the consumer group is consuming.

Every job in Liquid is divided into some tasks which consume different partitions of a topic for parallel processing. As a result, the maximum number of tasks in every job is limited to the number of partitions of a topic. Task number limitation violates the isolation of processing layer and messaging layer and limits the scalability of the jobs.

\begin{figure}[!t]
\begin{center}
\includegraphics[width=3in]{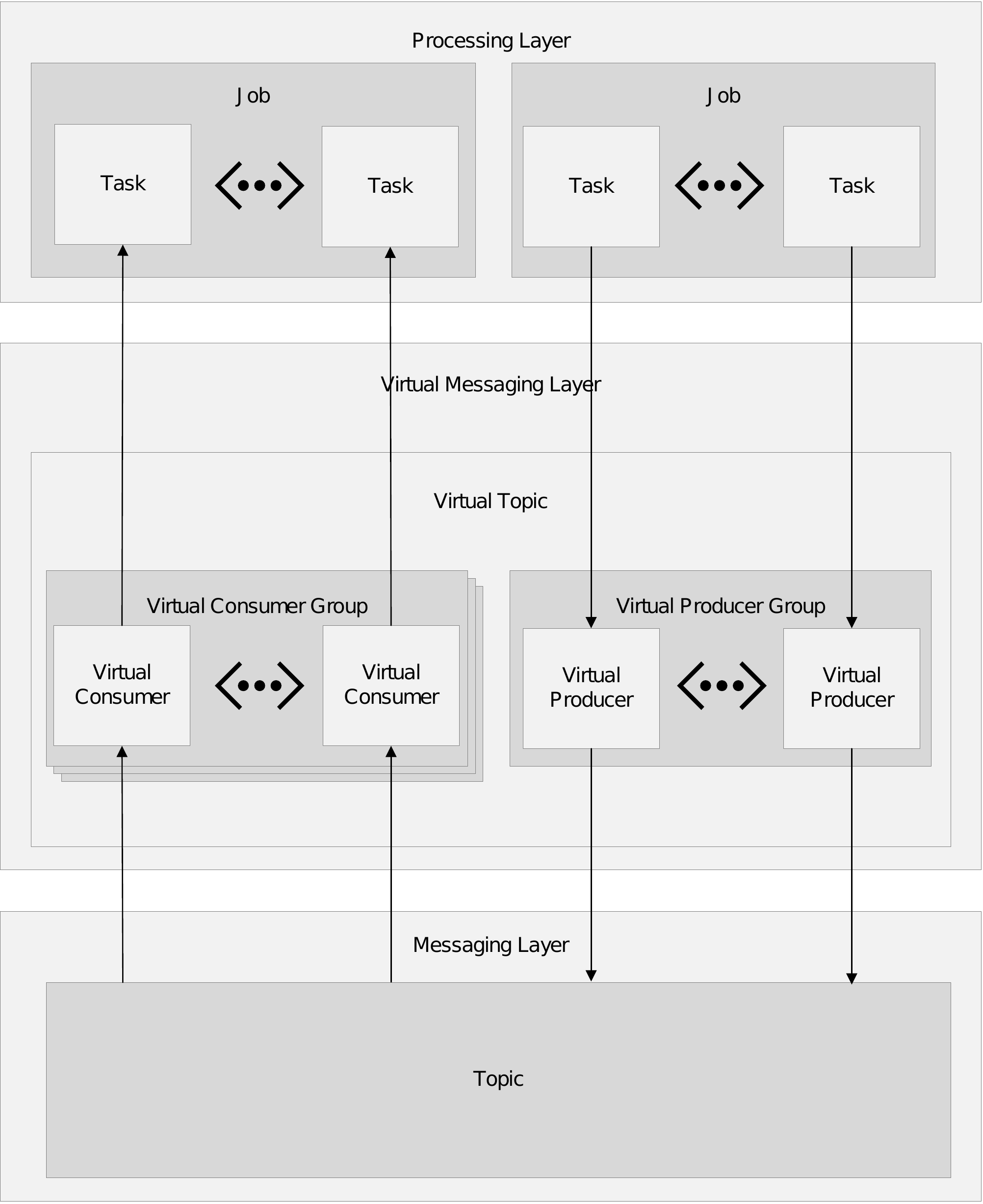}
\end{center}
\caption{The virtual messaging layer mediates between the processing layer and the messaging layer.}
\label{liquid_justification}
\end{figure}

To solve this problem, we propose a new layer between the messaging layer and the processing layer called the virtual messaging layer. This layer is shown in \figurename~\ref{liquid_justification}. There is a virtual topic in the virtual messaging layer corresponding to each topic in the messaging layer. The responsibility of a virtual topic is to consume messages from the messaging layer then distribute them among the tasks of the jobs which has subscribed to the topic. Furthermore, tasks can publish messages to the messaging layer through the virtual topics. Each virtual topic consists of a virtual producer group and zero or more virtual consumer groups.

The virtual producer group receives the messages which the tasks want to publish and distribute them among some producers. In fact, these producers publish the messages on the messaging layer, and the virtual producer group tries to balance the load of messages on producers.

Each virtual consuming group in a topic is created for a job which subscribes to that topic, and each one consists of one or more consumer. The consumers consume the messages from the messaging layer and distribute them among the tasks of the job. In the virtual messaging layer, the number of the consumers in a virtual consumer group is limited to the number of the corresponding topic. These consumers are usually faster than the tasks in a job because consuming a message and send it to a task is usually much simpler than processing a message. As a result, the tasks in a job can scale out corresponding to load of the incoming messages without being limited. In other words, the number of the tasks in a job are independent of the number of the partitions in the topic on which the job has subscribed.

\subsection{Reactive Liquid}
\begin{figure}[!t]
\begin{center}
\includegraphics[width=3in]{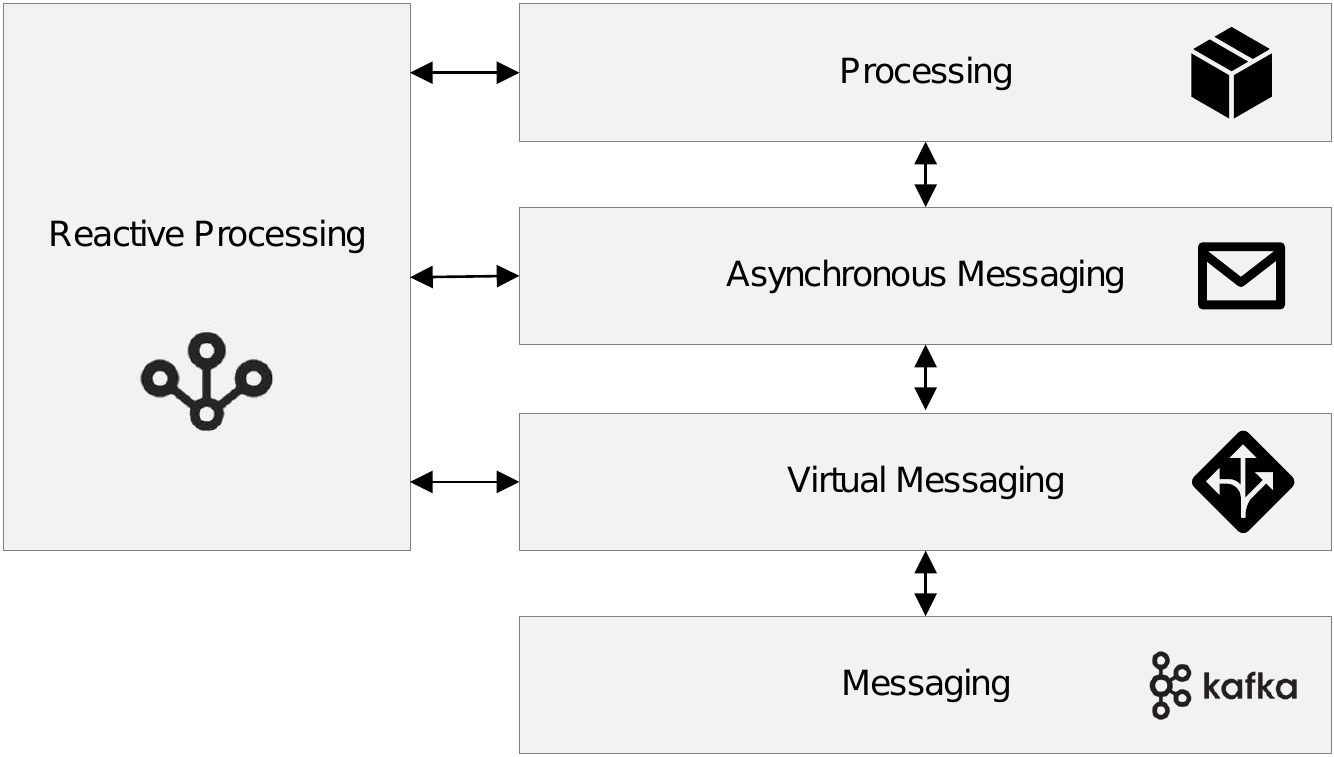}
\end{center}
\caption{The Reactive Liquid architecture}
\label{reactive_liquid_arch}
\end{figure}

We propose the Reactive Liquid as a highly scalable and resilient architecture to meet the requirements of existing distributed data processing systems. This architecture is shown in \figurename~\ref{reactive_liquid_arch}. Reactive Liquid is designed based on the reactive manifesto which makes the architecture responsive, resilient, elastic, and message-driven to that the architecture consists of five layers:

\begin{enumerate}
\item The messaging layer contains all data which are inputted into the system or produced by the jobs.
\item The reactive processing layer provides an elastic and resilient processing platform for the processing layer and the virtual messaging layer.
\item The virtual messaging layer provides a mediator between the messaging layer and the processing layer as was discussed in the preceding section.
\item The asynchronous messaging layer provides an asynchronous message-passing infrastructure to gain isolation and loose-coupling between the processing layer and the virtual messaging layer.
\item The processing layer runs the jobs and tasks in a distributed manner.
\end{enumerate}

\subsubsection{Messaging Layer}
This layer is the same messaging layer in original Liquid architecture which is responsible for data access. Messaging layer is a topic-based publish/subscribe platform which is highly available and scalable such as Apache Kafka.

\subsubsection{Reactive Processing Layer}
\begin{figure}[!t]
\begin{center}
\includegraphics[width=3in]{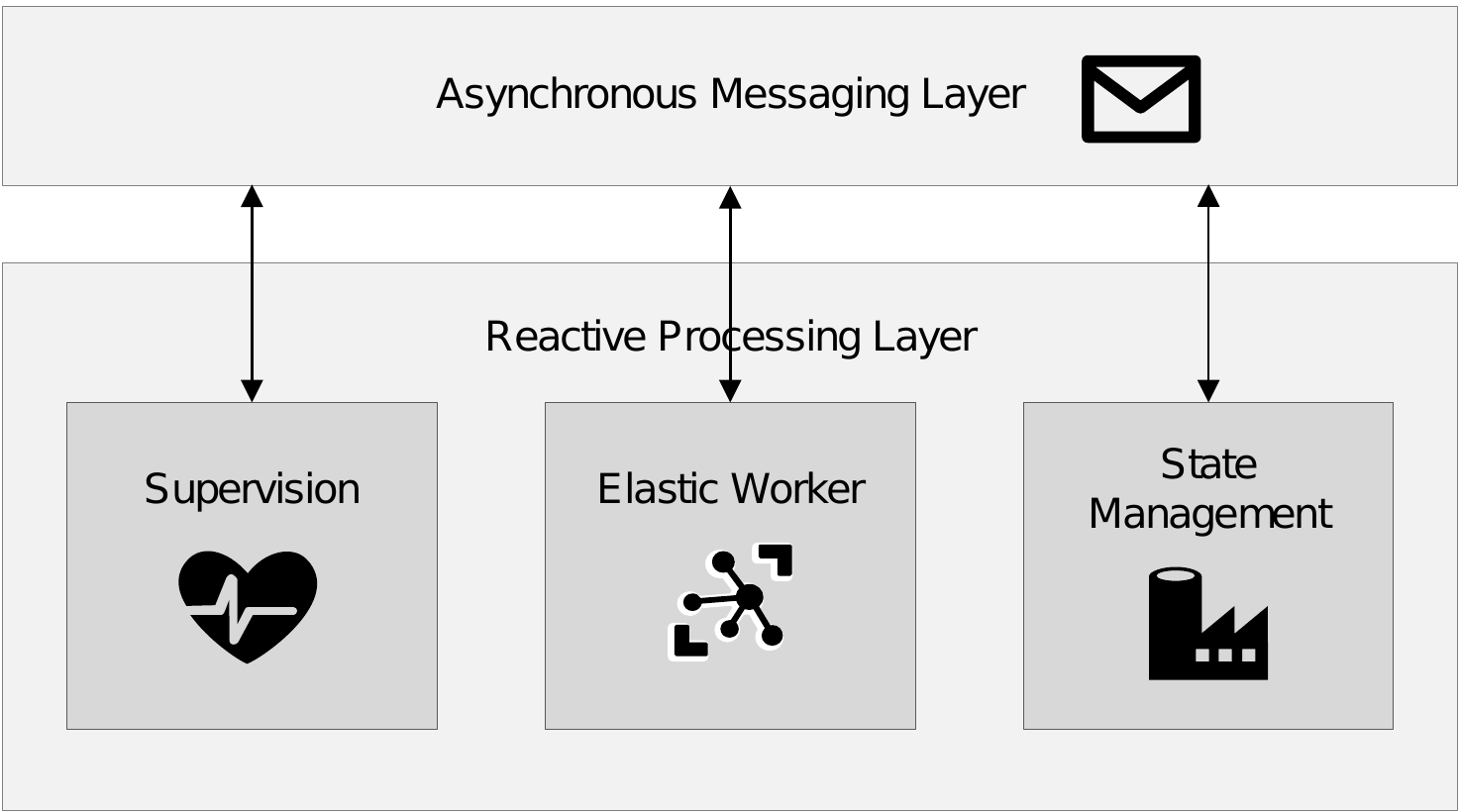}
\end{center}
\caption{Reactive Processing Layer}
\label{reactive_processing_layer}
\end{figure}

As shown in \figurename~\ref{reactive_processing_layer}, this layer provides three services as a reactive processing platform for the processing layer and virtual messaging layer. These services are Elastic Worker, Supervision, and State Management.

Elastic worker service is a dynamic scaling platform to provides auto-scaling workers which react to input workload and scale on demand. The elastic worker service monitors the message queue of the workers to estimate the workload. When the workload exceeds the agreed upper and lower limit, the service changes the number of the instances to fits the workload.

Supervision service applies the supervision pattern which was discussed in section 2. Supervision service provides a supervisor for essential components to check the health of those components and restart them if they fail.

State management provides a service for handling the state of stateful components. As mentioned in section 2, the reactive system should be able to restart its components. Thus, Reactive Liquid architecture should handle the persistent state of stateful jobs to enable the jobs to retrieve their state after restarting. Furthermore, the state management enables the distributed components to share their state without bottlenecks or contention points. The state management service provides persistent and immutable state by employing Event Sourcing Pattern which stores all changes to the state of a component as a sequence of events. Hence, other components can query these events without violating the component isolation also component can use the events to reconstruct its state \cite{Fowler2005}. Furthermore, the state management service uses CRDT, which is the short form of Conflict-free Replicated Data Type, to share the state between multiple distributed instances of a component. CRDT is a data structure which replicates data under eventual consistency across multiple distributed components in a network. Moreover, not only can system update the replicas independently and concurrently without coordination between them but also resolve inconsistencies mathematically \cite{Shapiro2011}.

\subsubsection{Virtual Messaging Layer}
\begin{figure}[!t]
\begin{center}
\includegraphics[width=3in]{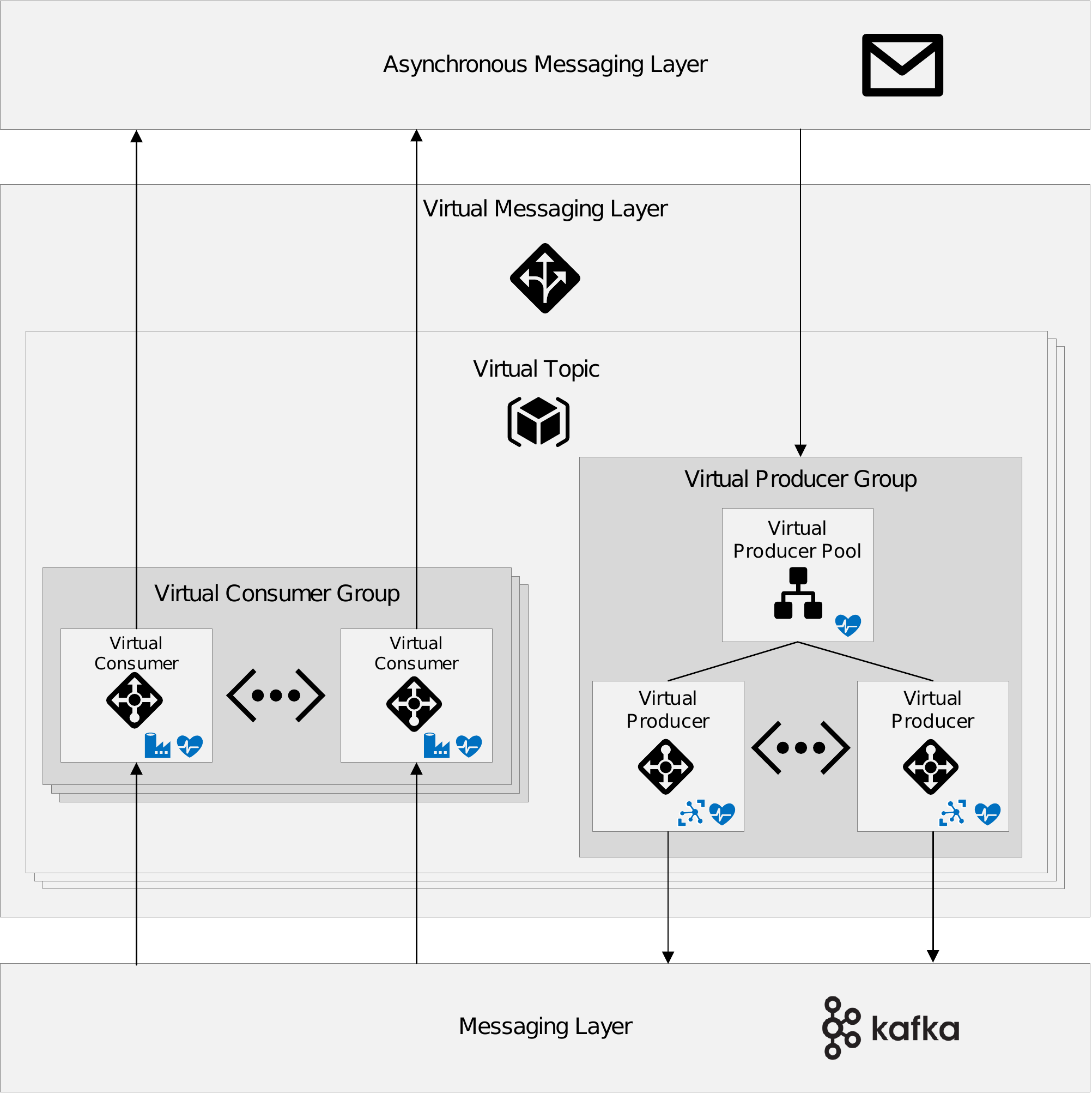}
\end{center}
\caption{Virtual Messaging Layer}
\label{virtual_messaging_layer}
\end{figure}

The virtual messaging layer obviates the scalability problems of Liquid Architecture by separating the consumer role from processing role. As shown in \figurename~\ref{virtual_messaging_layer}, virtual consumer groups of a virtual topic consume the messages from the corresponding topic in the messaging layer. Subsequently, virtual consumers in the virtual consumer groups send the consumed messages to the processing layer through the asynchronous messaging layer. Furthermore, virtual consumers and virtual producers are supervised by supervision service. If these components fail, their supervisors will restart them. Virtual consumers are stateful worker which persists the offset of the last consumed message. As a result, they can start consuming where they were stopped in case of a failure. The virtual producers use the elastic worker service to react to the incoming messages from jobs. Hence, the number of virtual producers depends on the incoming workload of the virtual topic. Virtual producer pool is responsible for distributing the messages and balancing the load among the virtual producers.

\subsubsection{Asynchronous Messaging Layer}
The asynchronous messaging layer provides asynchronous message-driven communication between the processing layer, virtual messaging layer, and reactive processing layer. In fact, asynchronous messaging layer makes the system message-driven and provides loose coupling, isolation, and location transparency among components as stated in the reactive manifesto.

\subsubsection{Processing Layer}
\begin{figure}[!t]
\begin{center}
\includegraphics[width=2.8in]{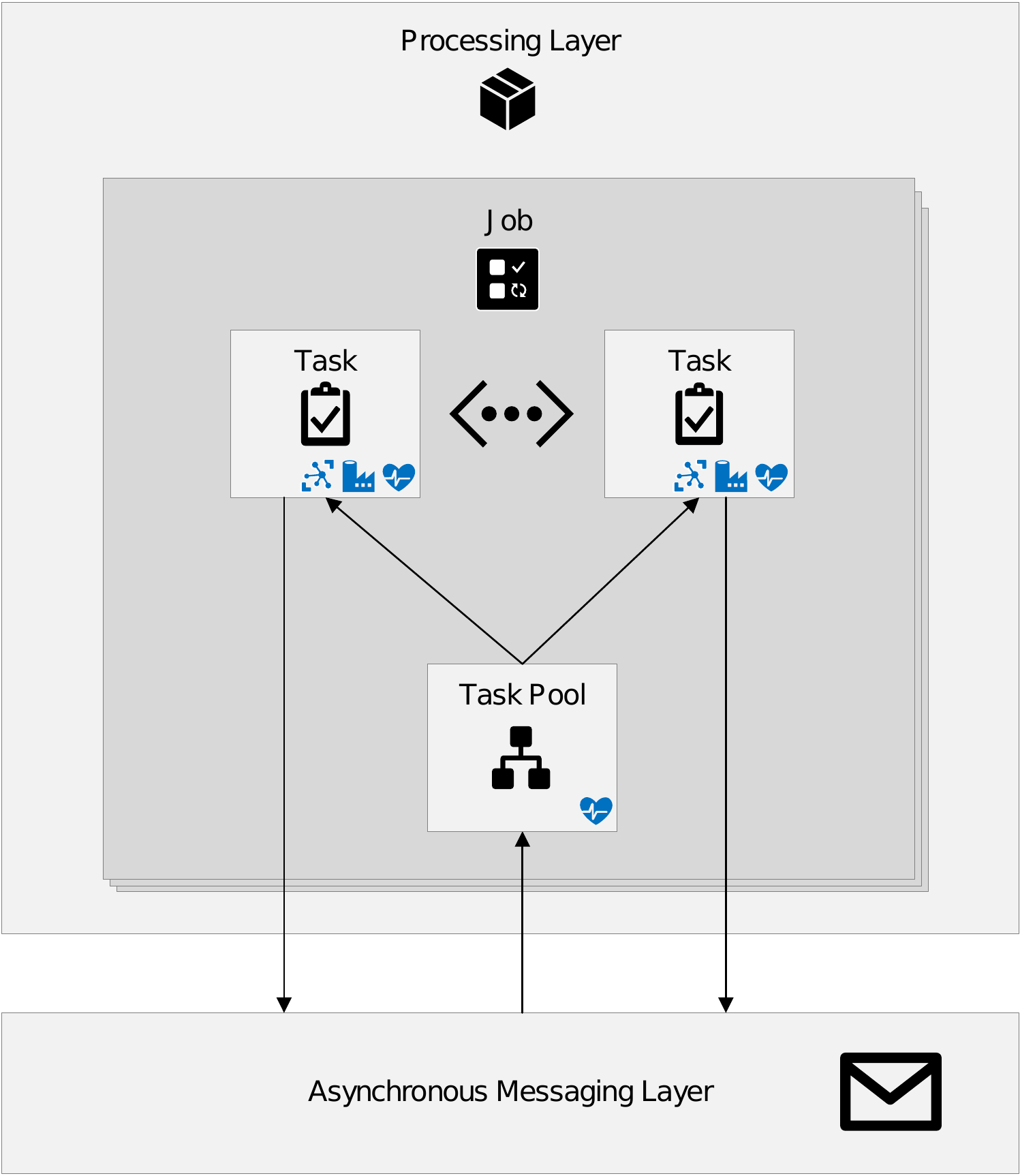}
\end{center}
\caption{Processing Layer}
\label{processing_layer}
\end{figure}

The processing layer distributes and runs the jobs across the network. As shown in \figurename~\ref{processing_layer}, every job consists of a number of tasks, which is based on the workload of the job. In other words, jobs employ elastic worker service to scale on demand. Moreover, stateful jobs use the state management service to persist the state and share the state among the tasks. Task pool distributes the messages and balances the load among the tasks of a job. Thus, the tasks will not compete for messages or be overloaded by incoming messages.

\section{Evaluation}\label{sec:evaluation}
\subsection{Implementation}
To evaluate the proposed Reactive Liquid architecture, we compare it with the Liquid architecture \cite{Fernandez2015}. We implement the messaging layer of the Liquid architecture by Apache Kafka and the processing layer by pure Java. Afterward, we construct the messaging layer of the Reactive Liquid architecture by Apache Kafka, the virtual messaging layer by pure Java, the asynchronous messaging by Akka, the processing layer by pure Java, and the reactive processing by Akka and pure Java.

To conduct the experiment, we implemente an incremental clustering algorithm for trajectories entitled TCMM\cite{Li2010}. TCMM breaks the clustering into two incremental steps, namely Micro- and Macro-clustering. First, micro-clustering step clusters the data to some micro-clusters which are defined as a temporal extension of the cluster feature vector. Furthermore, every data will merge with nearest existing micro-cluster or form a new micro-cluster. Next, a clustering algorithm will be applied on the micro-clusters to construct the macro-clusters in macro-clustering step. Macro-clustering step is executed periodically to keep abreast of the latest macro-clusters which are the evolving result of the algorithms. We create a micro-clustering job and a macro-clustering job corresponding to the steps of the TCMM. Moreover, the micro-clustering job consumes trajectories as messages from a topic in Apache Kafka and publishes the micro-clusters changes as an event source to a topic in Apache Kafka. Afterward, the micro-clustering job consumes the micro-clusters changes and publishes the macro-clusters changes as an event source to a topic too.

\subsection{Dataset}
The evaluation of the architecture is carried out on the taxi trajectories which is used by Jing Yuan et al. to mine driving directions from the GPS trajectories of 10,357 taxis \cite{Yuan2010}. This dataset was generated from February 2 to February 8, 2008, in Beijing city to evaluate cloud-based system which finds the customized and practically fastest driving route to a destination \cite{Yuan2011}.

\subsection{Experiment Setup}
We prepare a distributed environment consisted of 3 computing nodes which are equipped with 1.5 GB of RAM, 20 GB of hard disk, and a Core i3 Dual Core 2 GHz CPU. To perform the experiment, we prepare one implementation of the Reactive Liquid architecture and two implementations of the Liquid architecture. An implementation in which every job has precisely three tasks and another one with precisely six tasks. Furthermore, every topic of Apache Kafka in the messaging layer has three partitions in all of the implementations.

We compare the results of the implementations under four circumstances: every node fails after every 10 minutes working with a probability of zero percent, 30 percent, 60 percent, and 90 percent. Furthermore, every failed node restarts after 5 minutes. In every circumstance, we monitor the throughput of the system, the number of messages which the system process per second; the total number of processed messages every second from the beginning; the completion time of every message, the time when a message is consumed from messaging layer until it is entirely processed in processing layer.

\subsection{Evaluation Results}
\subsubsection{Without the probability of failure}
\begin{figure}[!t]
\begin{center}
\includegraphics[width=3in]{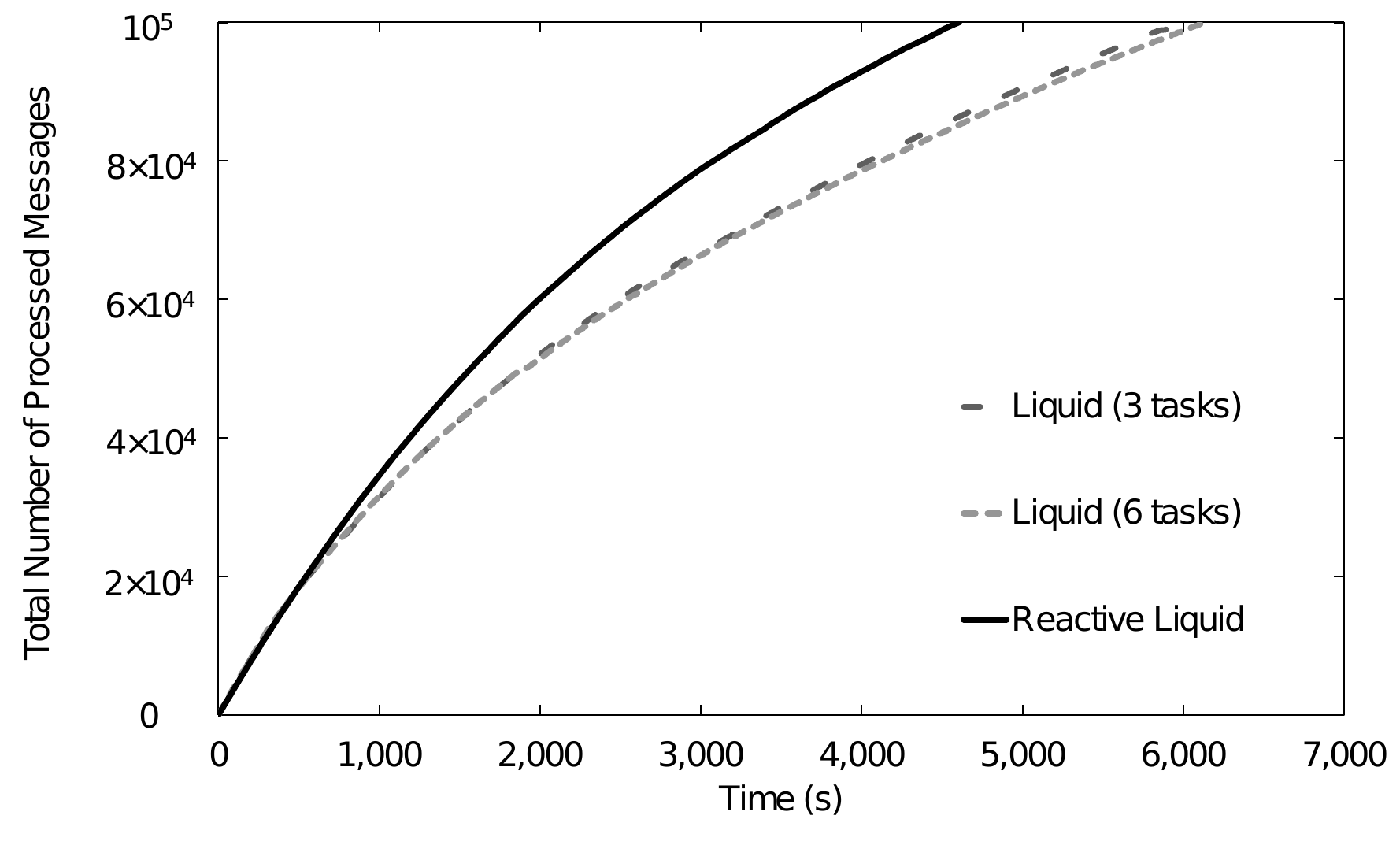}
\end{center}
\caption{The total number of processed messages for the Reactive Liquid and the Liquid implementations with three tasks and six tasks.}
\label{total_messages_no_failure}
\end{figure}

As mentioned before, every topic has three partitions. Hence, only three consumers can be active in every consumer group. Without the probability of failure, \figurename~\ref{total_messages_no_failure} shows almost the same total number of processed messages at every point in time for the Liquid implementations with both three tasks and six tasks. Nevertheless, the Reactive Liquid implementation has a higher total number of processed messages at every point in time than the Liquid implementations have since the limitation of the tasks is resolved in the Reactive Liquid architecture.

TCMM algorithm searches through the micro-clusters for the nearest one to input data. The micro-clusters size grows over time and decelerates the micro-clustering. Consequently, there is a gradual decrease in the slope of the curves for all implementations.

\begin{figure}[!t]
\centering
\subfloat[]{\includegraphics[width=2.9in]{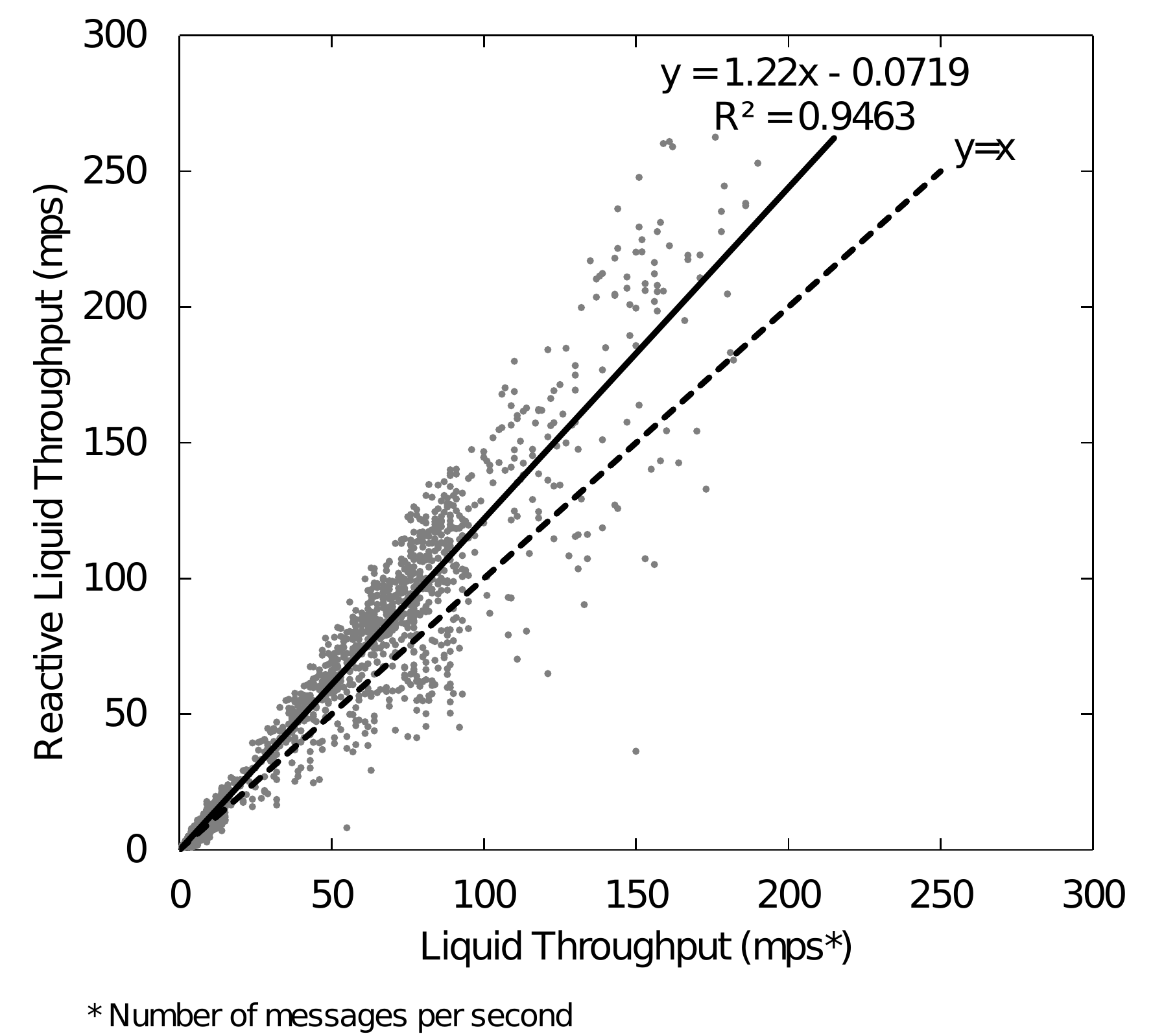}
\label{throughput_no_failure_first_case}}
\hfil
\subfloat[]{\includegraphics[width=2.9in]{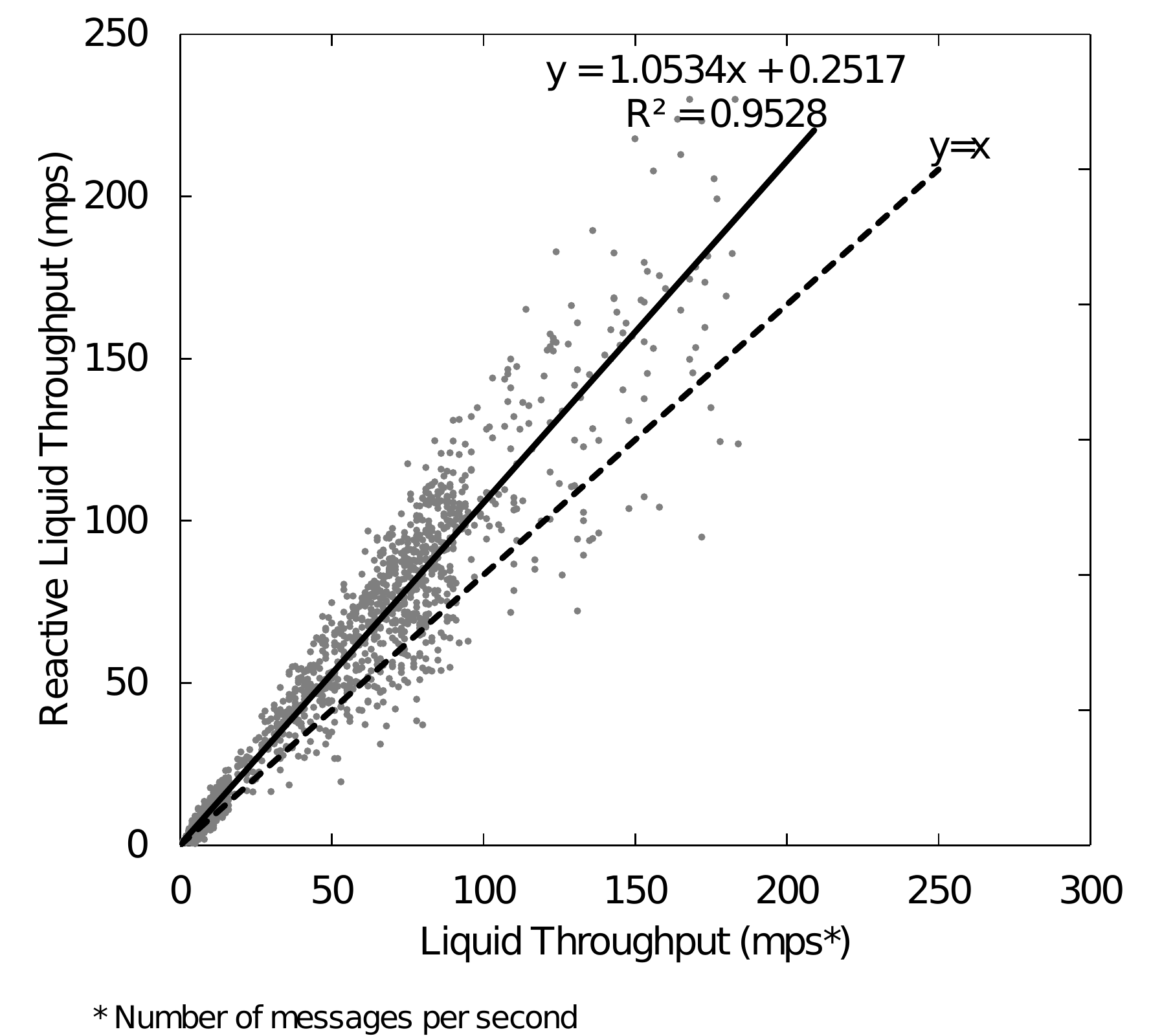}
\label{throughput_no_failure_second_case}}
\caption{(a) Comparison of throughput between the Liquid implementation with three tasks and the Reactive Liquid implementation. (b) Comparison of throughput between the Liquid implementation with six tasks and the Reactive Liquid implementation. The results show that the Reactive Liquid throughput is generally more than the Liquid throughput.}
\label{throughput_no_failure}
\end{figure}

\figurename~\ref{throughput_no_failure} draws a comparison of throughput between the Liquid implementations and the Reactive Liquid implementation. Every dot in the chart represents the number of processed messages of the Liquid implementation with three tasks compared to the Liquid implementation at a specified time. The linear trendline in the chart shows the overall direction of the data. Therefore, the position of this trendline compared to the y=x line depicts that which implementation has a better throughput generally. As is shown, the throughput of the Reactive Liquid is higher than the Liquid implementations.

The coefficient of determination known as R-squared assesses the goodness of fit of the trendline. Therefore, we used the R-squared to measure the accuracy of our comparison. R-squared values closer to 1 represent more perfect fits than the values closer to 0. As shown in \figurename~\ref{throughput_no_failure}, the R-squared measures of trendlines are higher than 0.9, which means the accuracy of our comparison is quite enough to rely on.

\subsubsection{With the probability of failure}
\begin{figure}[!t]
\centering
\subfloat[]{\includegraphics[width=3in]{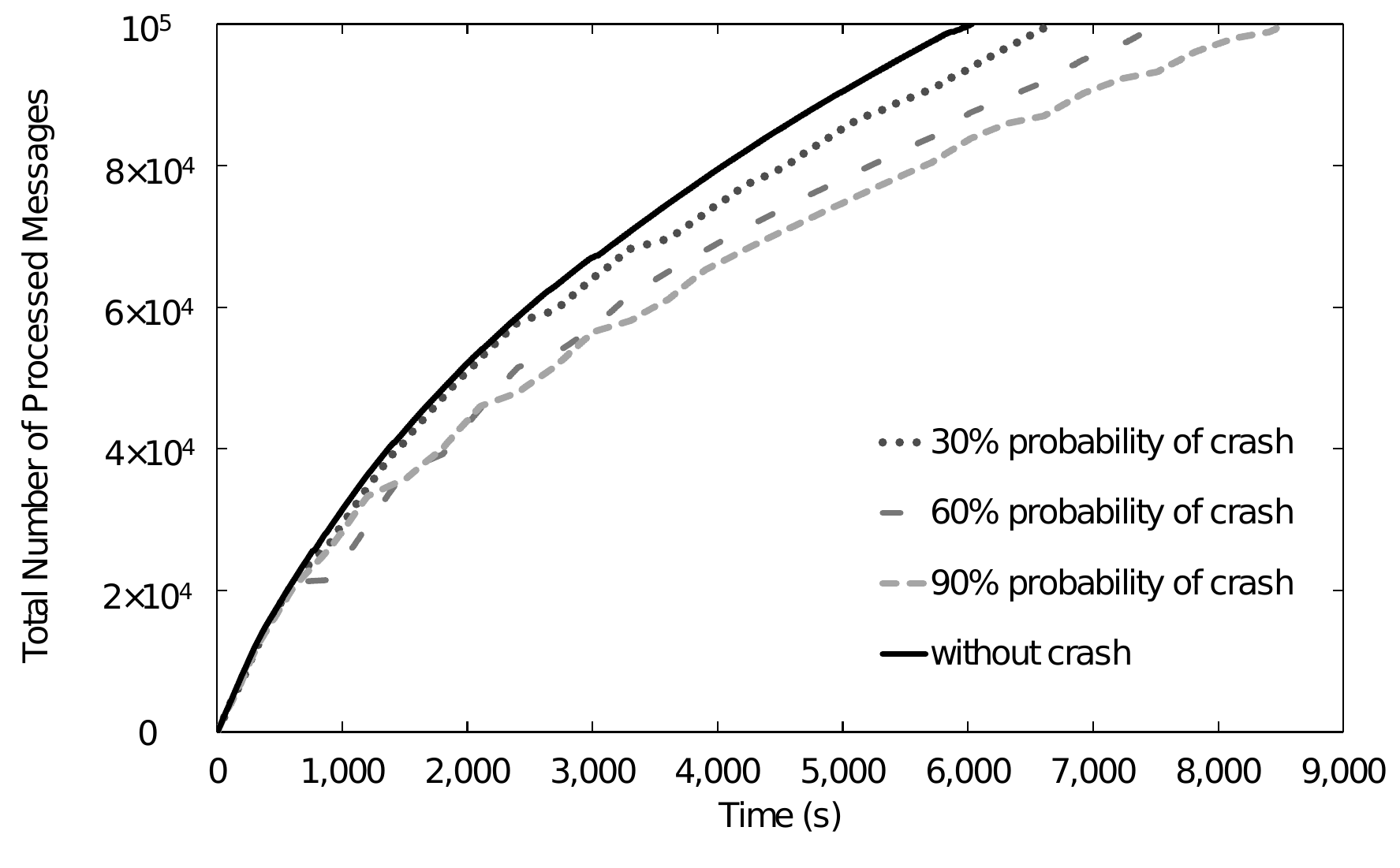}
\label{total_messages_with_failure_first_case}}
\hfil
\subfloat[]{\includegraphics[width=3in]{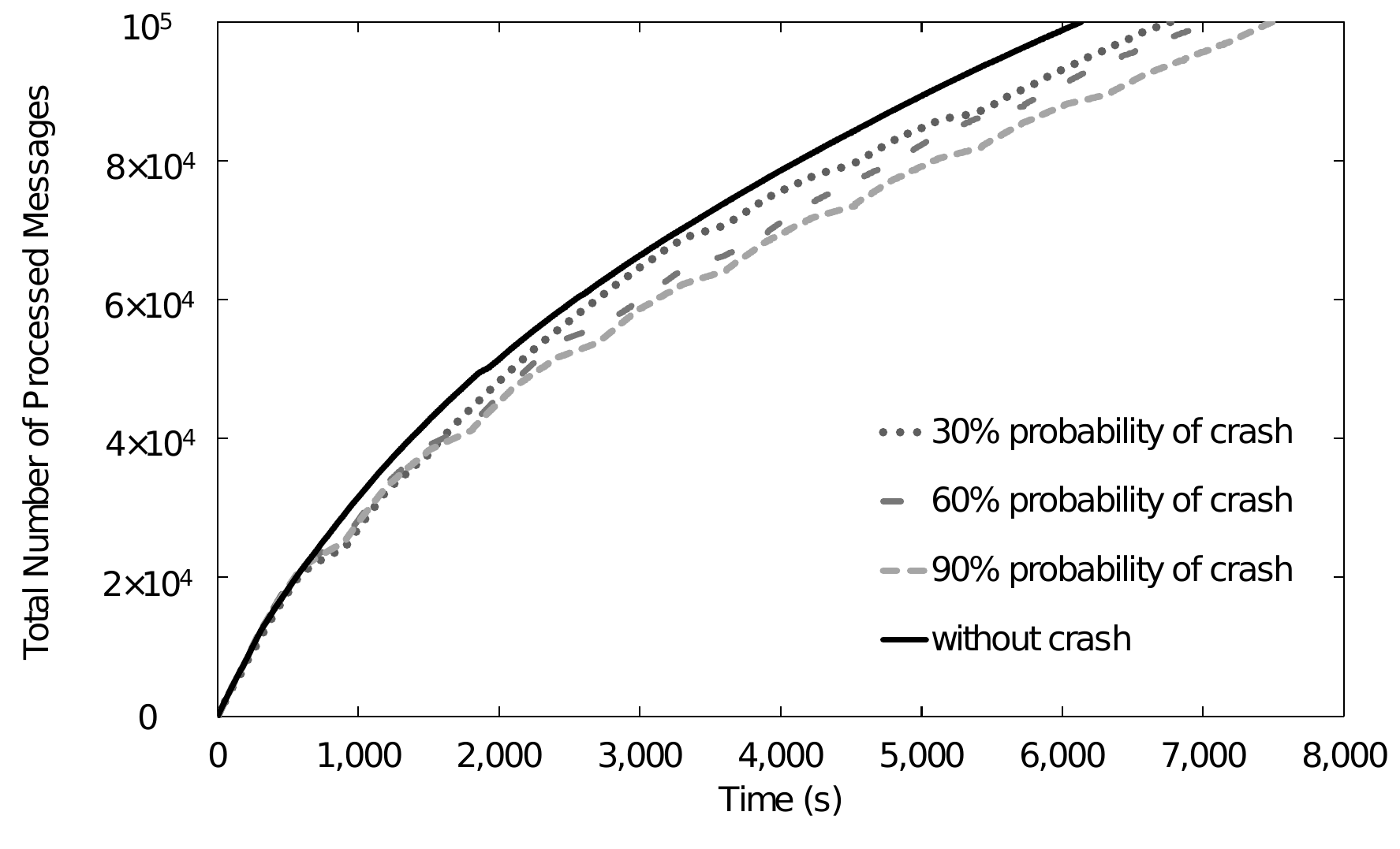}
\label{total_messages_with_failure_second_case}}
\hfil
\subfloat[]{\includegraphics[width=3in]{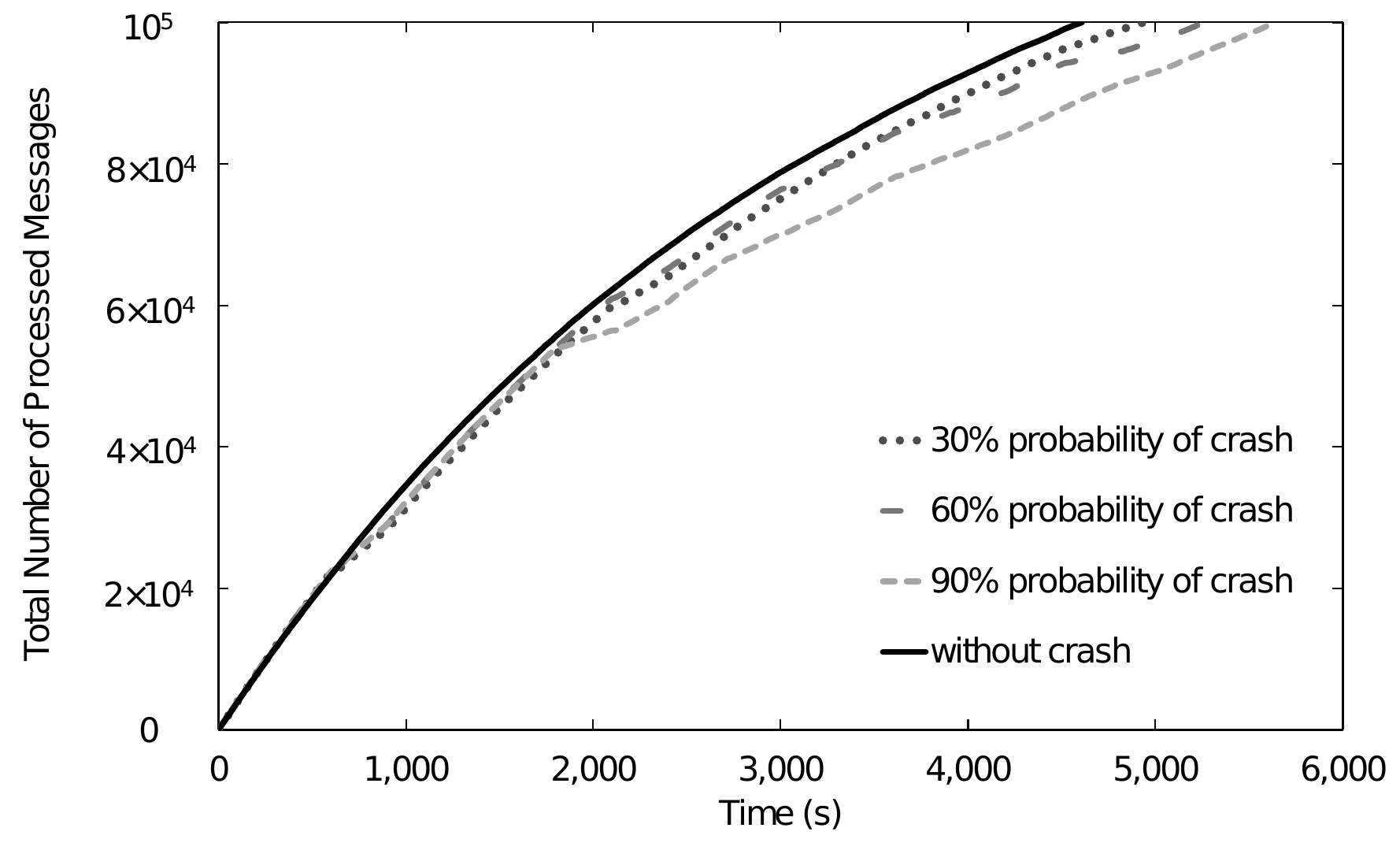}
\label{total_messages_with_failure_third_case}}
\caption{(a) The total number of processed messages for the Liquid implementation with three tasks with the probability of failure. (b) The total number of processed messages for the Liquid implementation with six tasks with the probability of failure. (c) The total number of processed messages for the Reactive Liquid implementation with the probability of failure.}
\label{total_messages_with_failure}
\end{figure}

In the case of a failure in a computing node, whereas the system loses some of the computing power, the supervision service of the Reactive Liquid detects the failed components and regenerates them in other healthy nodes, and the Reactive Liquid heals itself after an amount of time. \figurename~\ref{total_messages_with_failure} shows the effect of the probability of failure on the total number of processed messages in the Liquid and the Reactive Liquid implementations. The charts portray that the probability of failure affects the Liquid implementations more than the Reactive Liquid since the components of the Reactive Liquid are healed. The probability of failure affects the Reactive Liquid implementation too as not only does the computing power decrease but also the system takes time to detect the failure and heal itself.

\subsubsection{The processing time}
We measured the time when a message is consumed from the messaging layer until it is processed completely in the processing layer, which we refer to as the completion time of the message. In Liquid architecture every task consumes n messages, then it processes all n messages. After processing all of the consumed messages, the task consumes next n messages. Equation~\eqref{completion_time_liquid} represents a formula for calculating the completion time of the ith message where $t_c$ is the average consuming time of a message and $t_p$ is the average processing time of a message.

\begin{equation}
\label{completion_time_liquid}
T = n \times t_c+i \times t_p
\end{equation}

Every virtual consumer in Reactive Liquid architecture consumes n messages, then sends them to corresponding tasks and after that, it consumes next n messages. The formula for calculating the completion time of the ith message in Reactive Liquid is shown in Equation~\eqref{completion_time_reactive_liquid} where $t_c$, $t_p$, and $t_{wi}$ stand for average message consume time, average message process time, and ith message delay time (i.e the time the ith message takes to be sent to the corresponding task and waits in the task message queue) respectively.

\begin{equation}
\label{completion_time_reactive_liquid}
T = n \times t_c+t_{wi}+t_p
\end{equation}

\begin{figure}[!t]
\centering
\subfloat[]{\includegraphics[width=2.9in]{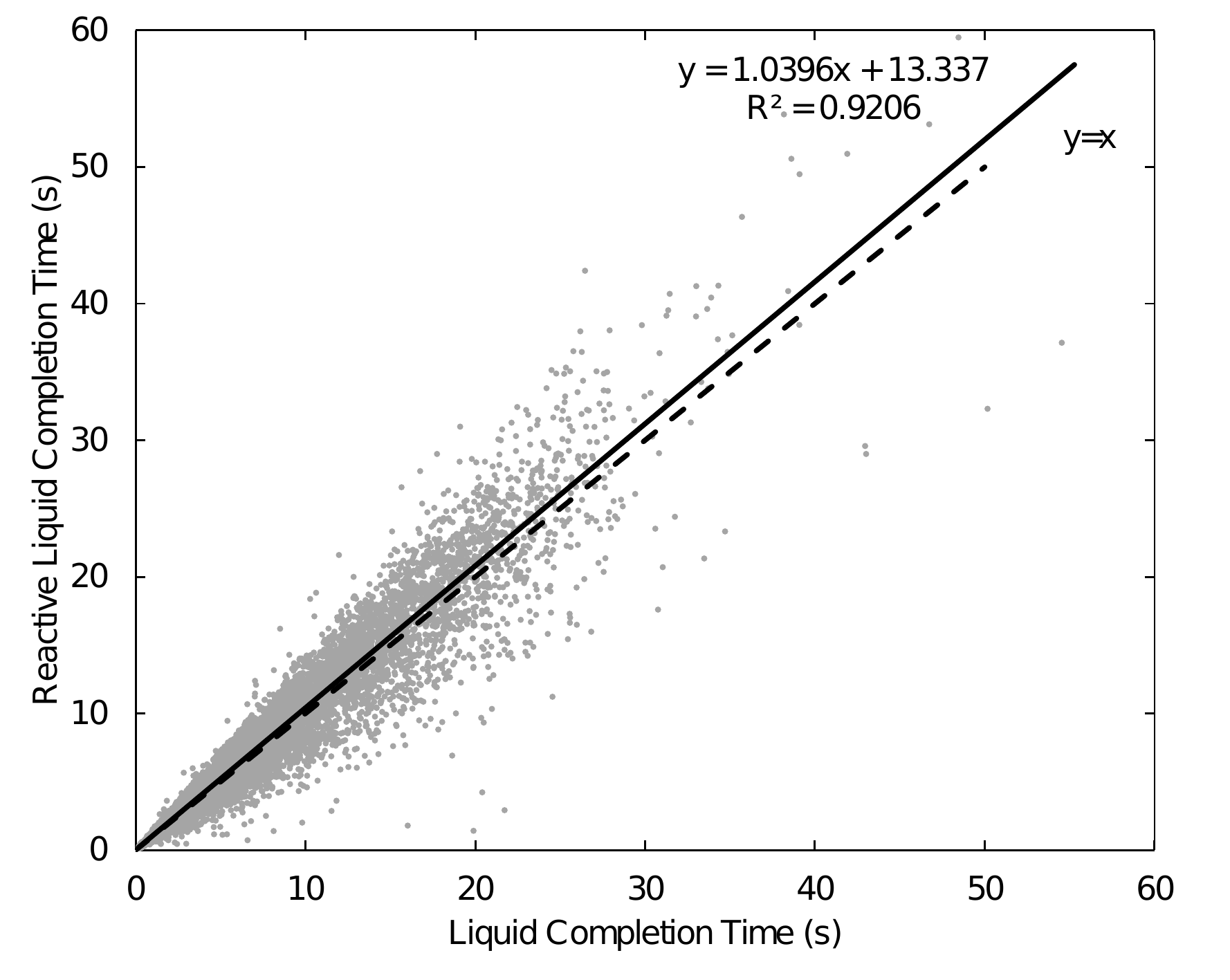}
\label{completion_time_first_case}}
\hfil
\subfloat[]{\includegraphics[width=2.9in]{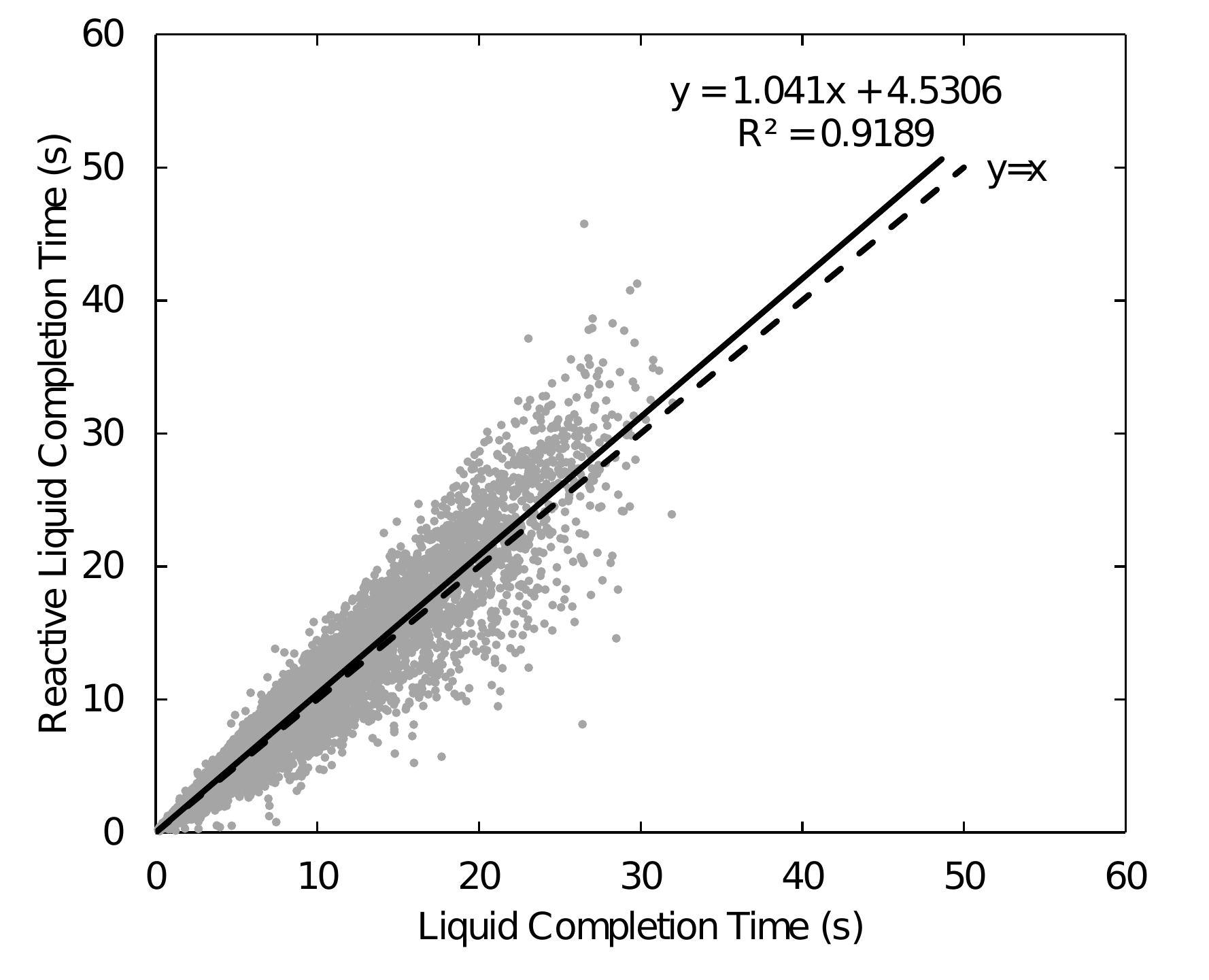}
\label{completion_time_second_case}}
\caption{(a) Comparison of completion time between the Liquid implementation with three tasks and the Reactive Liquid implementation. (b) Comparison of completion time between the Liquid implementation with six tasks and the Reactive Liquid implementation.}
\label{completion_time}
\end{figure}

The value of $t_{wi}$ depends on the size of the message queue and the number of the messages in the queue. In Reactive Liquid, every virtual consumer consumes and delegates messages successively without interruption, and the messages are usually consumed much more quickly than they are processed. Consequently, the values of $t_{wi}$ are usually vast and unstable, which can lead to adverse completion time. \figurename~\ref{completion_time} shows that the completion time of the Reactive Liquid implementation is generally more than the Liquid implementation.

\section{Conclusion}
We presented Reactive Liquid, a highly scalable and resilient distributed architecture based on the Liquid architecture. The Reactive Liquid delivers the reactive manifesto promises which are responsiveness, resiliency, elasticity, and message-driven communication.

The Reactive Liquid detects the failed components as quickly as possible and regenerates them to heal the system. Moreover, The Reactive Liquid adjusts the usage of the resources to fit the computation power to input workload. In other words, the Reactive Liquid is elastic which means scalable on demand.

Our work leaves an open research problem. The completion time of the Reactive Liquid architecture is generally more than the Liquid architecture, which contradicts the reactive manifesto. Nevertheless, the Reactive Liquid architecture satisfies the requirements of a reactive system. It seems that the need for a message distribution scheduler algorithm which distributes the messages among the tasks is crucial to minimize the completion time of the messages.

\bibliographystyle{unsrt}  
\bibliography{references}
\end{document}